\documentclass[sn-mathphys-num]{sn-jnl}%

\usepackage{graphicx}%
\usepackage{multirow}%
\usepackage{amsmath,amssymb,amsfonts}%
\usepackage{amsthm}%
\usepackage{mathrsfs}%
\usepackage[title]{appendix}%
\usepackage{xcolor}%
\usepackage{textcomp}%
\usepackage{manyfoot}%
\usepackage{booktabs}%
\usepackage{algorithm}%
\usepackage{algorithmicx}%
\usepackage{algpseudocode}%
\usepackage{listings}%

\begin{document}

\title[Article Title]{A computational materials science paradigm for a Course-based Undergraduate Research Experience (CURE)}

\author[1,2]{\fnm{David A.} \sur{Strubbe}}\email{dstrubbe@ucmerced.edu}

\affil[1]{\orgdiv{Department of Physics}, \orgname{University of California, Merced}, \orgaddress{\street{5200 N. Lake Rd.}, \city{Merced}, \postcode{95343}, \state{CA}, \country{USA}}}

\affil[1]{\orgdiv{Materials and Biomaterials Science and Engineering Graduate Group}, \orgname{University of California, Merced}, \orgaddress{\street{5200 N. Lake Rd.}, \city{Merced}, \postcode{95343}, \state{CA}, \country{USA}}}

\abstract{
Course-based Undergraduate Research Experiences (CUREs) bring the excitement of research into the classroom to improve learning and the sense of belonging in the field. They can reach more students, earlier in their studies, than typical undergraduate research. Key aspects are: students learn and use research methods, give input into the project, generate new research data, and analyze it to draw conclusions that are not known beforehand. CUREs are common in other fields but have been rare in materials science and engineering. I propose a paradigm for computational material science CUREs, enabled by web-based simulation tools from nanoHUB.org that require minimal computational skills. After preparatory exercises, students each calculate part of a set of closely related materials, following a defined protocol to contribute to a novel class dataset which they analyze, and also calculate an additional property of their choice. This approach has been used successfully in several class projects.
}

\maketitle

\begin{figure}[h]
\centering
\includegraphics[scale=0.75]{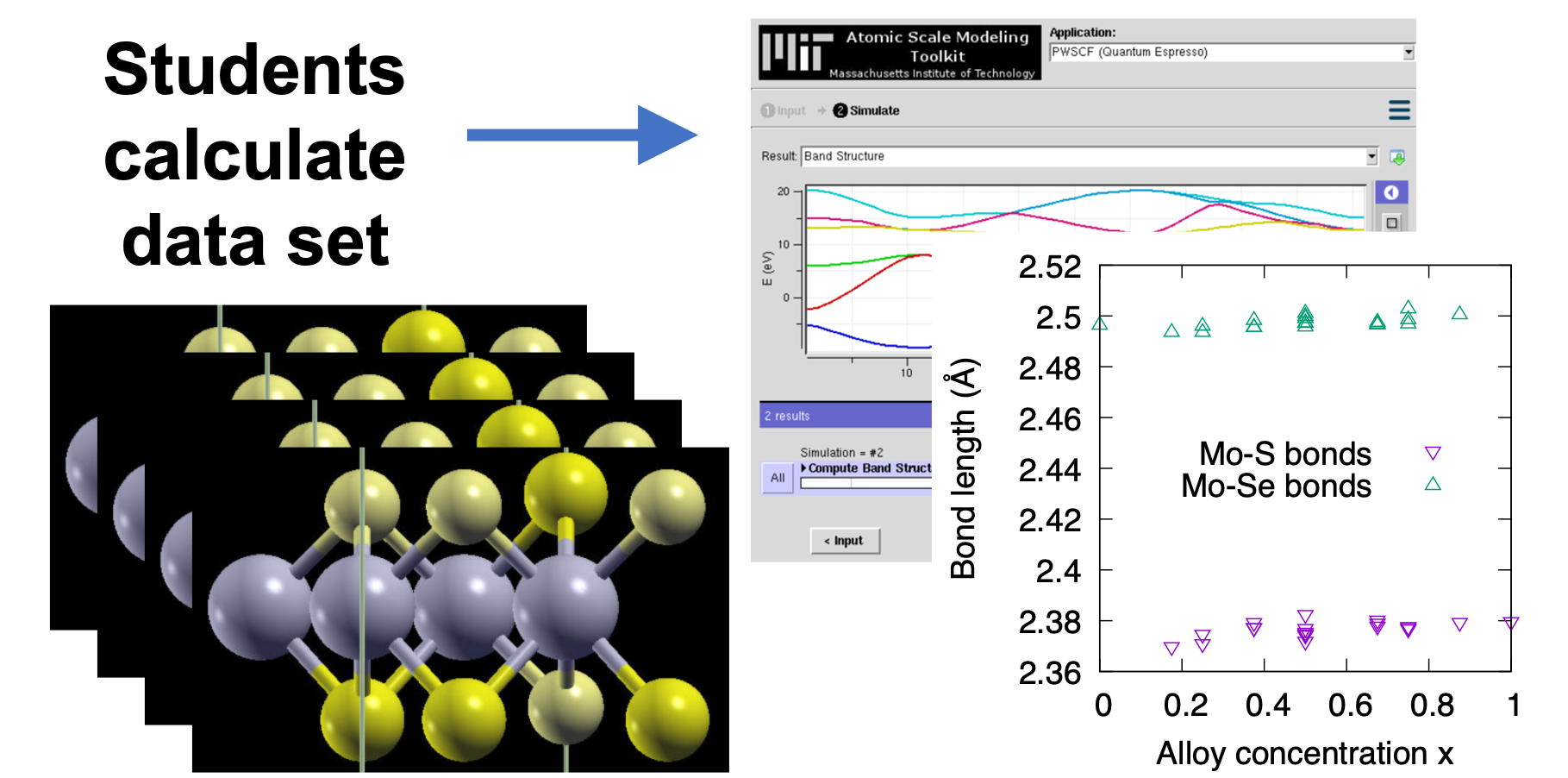}
\caption{Graphical abstract.}
\end{figure}

\section{Introduction}\label{sec1}

A Course-based Undergraduate Research Experience (CURE) is an educational paradigm that brings the excitement of research into the classroom \cite{Lopatto,Waterman}. Key aspects of a CURE, in contrast to traditional ``cookbook'' or verification-oriented laboratory exercises, are: students learning and using research methods, having input into the project, generating new research data, and analyzing it to draw conclusions that are not known beforehand \cite{Lopatto,Brownell}. Research studies have shown that the CURE paradigm improves learning and motivation, promotes independent thinking, and increases retention of students in the major and STEM study. In fact, a CURE can affect attitudes and motivation sometimes as much as a full summer research experience \cite{DiBartolo,Lopatto}. CUREs also give students an opportunity to apply their knowledge and get a taste of research. Instructors can make use of CURE best practices and how-to guides that have been developed \cite{Lopatto,Waterman}, and a clearinghouse of CURE activities at CUREnet \cite{CUREnet}. While CUREs have become popular in biology \cite{DiBartolo} and to a lesser extent chemistry \cite{Waterman}, they remain rare or absent in materials science and engineering, as well as condensed-matter physics. This article proposes a paradigm for CUREs using computational materials science, to enhance curricula and broaden participation in undergraduate research in the materials science and engineering community.

Many students who begin as science and engineering majors end up not continuing in the field after graduation, or even leaving the major \cite{PCAST}. This is particularly the case for under-represented minorities (URMs) or minoritized students, who often feel a lack of support \cite{TEAM-UP}. A proven approach to improve persistence and retention is to help students identify as scientists, feel a sense of belonging, and experience what real research is like \cite{DiBartolo, Lopatto}. Participation in a CURE brings the excitement of research into the classroom, and allows students to experience doing real science. This helps promote the sense of belonging to a STEM community of practice as a budding scientist, which is particularly important for URMs \cite{TEAM-UP, Lopatto}. Studies have found that CUREs are particularly beneficial for URMs \cite{DiBartolo,Malotky,Bangera}. Introduction to research in the CURE helps transition students to summer research opportunities and possible senior thesis research. Open-ended experimental projects incorporated into an introductory materials science class were found to improve students' knowledge gain and skills \cite{Zhou}, though -- unlike a CURE -- these research-like projects were not necessarily designed to create new knowledge.

Computation is a particularly suitable kind of research for a CURE. Computation can be used to bridge between the simple models commonly studied in introductory classes and more complicated real materials. It can be cheap, well-defined, and easily reproducible, which is not necessarily the case for experimental work. In the specific case of atomistic simulations in materials science, it is easy to get to the frontiers of knowledge: given the vast space of materials, even slight changes to the structure and composition almost certainly lead to something that has not been previously studied. CUREs have generally been wet-lab activities, and there are few computational examples (e.g. bioinformatics \cite{Dahlquist_Dionisio_Libeskind-Hadas_Bargagliotti_2018}). Some other group projects in computational materials, which may have some of the aspects of a CURE, have been run in a graduate condensed-matter summer school and an undergraduate/graduate chemistry class, resulting in research articles \cite{Hummelshoj} and \cite{Shumilov} respectively.

\begin{figure}[h]
\centering
\includegraphics[width=\textwidth]{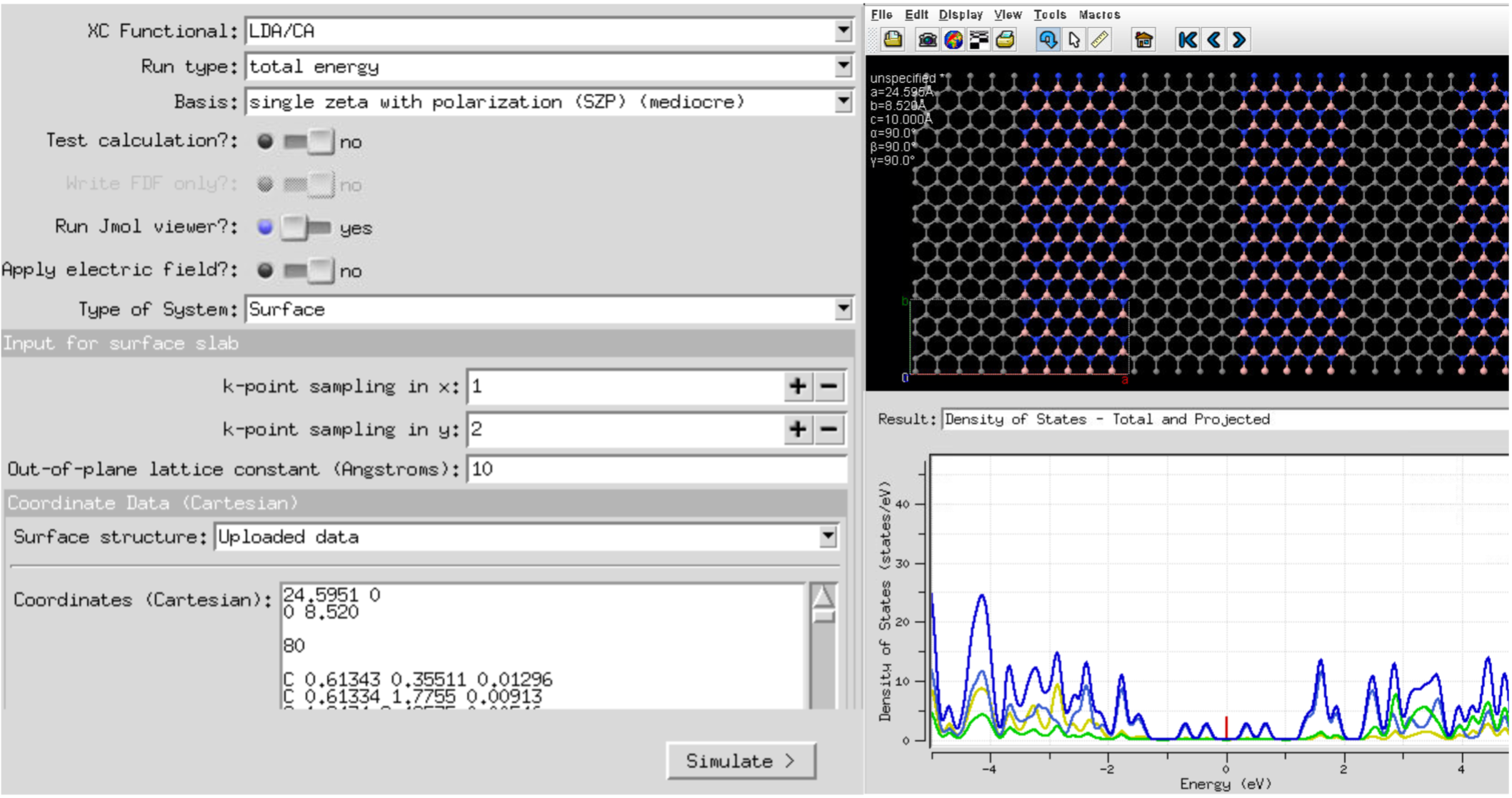}
\caption{Example calculation in nanoHUB graphical user interface of MIT Atomic-Scale Modeling Toolkit \cite{toolkit}: calculation of graphene/hBN superlattice heterostructure, by density-functional theory in Quantum ESPRESSO \cite{QE-2009}}\label{fig1}
\end{figure}

Computational work has become ubiquitous in research, but the typical workflow can feel quite daunting to undergraduates (Linux, compiling, running jobs, etc.). Excitingly, advances in codes and technology now enable some research-grade computation to be fast and accessible enough for an undergraduate course. The nanoHUB project \cite{nanoHUB} provides a convenient platform for real computations with simple graphical user interfaces (GUIs) that run in a web browser. The GUIs (``tools''), created for various codes by expert users, abstract away extraneous details that create a barrier, allowing direct engagement with the key ideas. The GUI creates input files based on student-specified simulation parameters, and then launches the calculations in the cloud on remote clusters, removing the need for any specialized or powerful hardware. After the run, the nanoHUB tools analyze output, and organize and plot key results (Fig. \ref{fig1}).  
While numerous courses use nanoHUB (e.g. \cite{Reeve}), I am not aware of use in a CURE other than my projects mentioned below.

In this article, I describe a paradigm for designing CUREs in computational materials science. Depending on instructional needs and capacity, a CURE can be focused on a single lab session or assignment, take the form of a culminating final project, or even constitute the entirety of a course. Students use nanoHUB to calculate each of a set of structures, forming a class dataset which they analyze along with their individual results. CUREs not only enhance the student experience but can also help instructors in their research and serve as compelling components on broader impacts or education for instructors' grant proposals.

\section{Materials and Methods}\label{sec11}

The nanoHUB project provides a large number of different tools that implement various kinds of simulations, including not only atomistic simulations but also electronic device, nanomechanical device, and biomolecular sensing simulations. The GUIs enable students to perform calculations without requiring deep knowledge of computational materials science. Here I will focus on the MIT Atomic-Scale Modeling Toolkit \cite{toolkit}, which I co-developed and which is used at many institutions around the world. It is based on open-source codes that are commonly used by researchers: it can run first-principles density-functional theory (DFT) calculations on materials with the pseudo-atomic orbital code SIESTA and the plane-wave code Quantum ESPRESSO \cite{QE-2009}; DFT and quantum chemistry on molecules with the Gaussian-type orbital code GAMESS \cite{GAMESS}; and classical molecular dynamics (MD) in the LAMMPS code \cite{LAMMPS} (e.g. for carbon nanostructures with Tersoff potentials). The DFT tools can provide relaxed structure, energy, density of states, band gap, bandstructure, wavefunctions, Kohn-Sham potential, electronic density, phonon bandstructure and density of states, Raman and infrared spectra, and phonon displacement patterns. MD can provide the relaxed structure or trajectories, radial distribution functions, and power spectra of vibrations. The toolkit integrates visualization with XCrySDen \cite{xcrysden} and Jmol codes, to study crystal and molecular structures, phonon displacement patterns, wavefunctions, densities, potentials, Fermi surfaces, and Brillouin zones.
The restricted input options in this GUI help in specifying a detailed protocol for students to use to ensure appropriate and systematic results in the resulting dataset. The author and his former PhD student Enrique Guerrero have presented a series of webinars \cite{webinar1, webinar2, webinar3, webinar4} for nanoHUB, with accompanying handouts to follow along in the simulations \cite{nanohub_handouts}, which can help to inform use of the tool and how it can be employed for a CURE.

The key steps of this computational CURE paradigm are laid out in Table \ref{tab1}. To design a CURE, the instructor should choose a research question that can be answered by students' calculations with the toolkit, and which will involve desired calculation methods or course concepts. The instructor selects a set of materials structures, which will be divided up among the students (e.g. one each). Structures can be selected using databases such as the Materials Project \cite{MaterialsProject}, and results can even be contributed back to the database. Generally it is not feasible for the students to select the research question and materials, because deep expert knowledge is required to design an authentically novel research study.
Some ideas for generating a suitable materials space include alloys (enumerating symmetry-unique structures within a certain supercell for different concentrations), dopants, polymorphs, surfaces, and interfaces or heterojunctions between different materials. Each of these allows a combinatorial construction of a set of similar materials for the students to study.
The research question could involve trying to maximize or minimize some property (e.g. find which alloy structure is most stable), or the analysis of some global property of the data set (e.g. find how the band gap varies with alloy composition). To analyze the data, students apply their knowledge of course concepts, and could use analytical models (e.g. Vegard's Law) or potentially more sophisticated informatics or machine-learning approaches. The instructor (or research group members) should run one or more examples in detail to verify the workflow and be aware of potential problems, which may sometimes occur in only some items of the set -- two examples confronted in my CURE with Quantum ESPRESSO were (1) incomplete variable-cell relaxations with residual stress and (2) imaginary phonon frequencies that were later attributed to the pseudopotential. Of course, the only way to completely exclude such problems is to do all the students' calculations beforehand, which defeats the purpose of a CURE.

\begin{table}[h]
\caption{Phases of computational materials science CURE}\label{tab1}%
\begin{tabular}{@{}llll@{}}
\toprule
Phase & Main Goals\\
\midrule
Lecture(s) & Introduction to concept of a CURE, background and motivation for \\
& research, methods and concepts to be used, practical steps in project  \\
Discussion section exercises & Familiarization with nanoHUB environment and the use of \\
& specific nanoHUB tools \\
Homework assignments & Practice in use of nanoHUB tools and perform preliminary steps in \\
& project, as checkpoint for instructor feedback \\
Submission of raw data & Collection of class dataset for students to analyze \\
Final presentation/paper & Analysis of individual data and class dataset, exploration of individual choice of \\
& property to calculate or investigate \\
\botrule
\end{tabular}
\end{table}

The CURE begins with one or more lectures which introduce the research question, show what remains unknown, and indicate how the CURE studies fit into a cutting-edge research project. Theory and needed concepts for running the code and analyzing results are also explained. Plans can be described for following up interesting findings with further calculations and/or experiments. In some cases, teaching-assistant training may also be needed. A detailed lab manual with instructions (e.g. \cite{lab_manual}) is given to the students. Ideally, the lab manual will be tested with
undergraduates before use in class to identify and improve on points of confusion or technical problems which can occur. In some cases, I have also engaged in further development of the MIT Atomic-Scale Modeling Toolkit to provide new functionality needed for the CURE. I will focus on the use of a CURE as a final project, which can be the most impactful to deploy within an existing course, since it enables students to have an extensive preparation period and then spend substantial time on the project. In-class demonstrations can introduce the usage of nanoHUB. In discussion sections, students engage in nanoHUB exercises themed with the recent lecture materials, serving both to illustrate and deepen their understanding of the lecture concepts but also to gain experience with the tools. They can further apply this knowledge in homework assignments, for more detailed feedback from the instructor. It is highly recommended to have some key early stages of the project be carried out in homework (e.g. structural relaxation, before calculation of other properties later) to ensure that the instructor can point out and correct any potentially serious problems in students' calculations at an early stage.

Then the students carry out their main calculations, following a carefully defined protocol, so that their results constitute a set of comparable data that the class can analyze. They submit their key results such as bandgaps to a repository (e.g. a common spreadsheet). It should be emphasized to students that when they analyze this data they should keep a close eye out for any anomalies in the data, which can be real physically interesting phenomena, artefacts due to sophisticated computational problems, or the result of trivial cut-and-paste mistakes. Students should submit not only key results but also all their raw input and output (which are downloadable from the tools in the MIT Atomic-Scale Modeling Toolkit, and many of the other nanoHUB tools) to a shared folder to make it possible for the instructor or other students to audit the calculations, and trace any errors or mistakes. As much as possible, students should identify and fix any anomalies. 

The need for the defined protocol in creating the dataset should be balanced with a way of allowing the students to have input and agency in the project (one of the key aspects of a CURE), to be sure to give a flavor of the creative and open-ended aspects of research. I have done this in two ways: in a short CURE consisting of just a single lab session, students dig into further outputs and information provided by the code beyond those they were instructed to study by the lab manual, and are asked to report something interesting they found. Quite lively discussions can result in the lab session as they try to make sense of what they found and understand its significance. In a final-project CURE, instead I ask students, in an exploratory phase of the project, to identify another property to calculate and analyze. I offer some examples, and they should read a bit about a property and discuss with the instructor how they can extract such information. Calculations of response properties like elastic moduli by finite differences are a good example. To enable such exploratory phases, it is important that the tool should be rich enough to provide more information than the minimum of what is required in the project. Students then present or write in the final project about the meaning of their property, their approach to calculate it, and their results. Note that typically it would require more student knowledge than is attainable in the context of the course to make these additional property calculations constitute reliable research data.

After the CURE, there can be follow-up by members of the instructor's research group (ideally even a student from the class as a summer researcher or postbac researcher), to confirm or refine the calculations and prepare them for publication or further directions of calculations or experiments. Such students can also work on improving the GUI and lab activity, or preparing to take the research question in a new direction for a coming year.

Improvement or adjustment of the CURE after each class should be informed by assessment. Enhanced learning of the class material can be assessed by analysis of students' class work.
To assess the effect on students’ attitudes and scientific thinking, two research-validated surveys are available: the Colorado Learning Attitudes about Science Survey (CLASS), a pre-/post-survey commonly used in introductory classes \cite{CLASS}; and the ``CURE survey'' \cite{CURE_survey}, which includes assessing the feeling of community. The instructor's observations of students at work in lab and discussion sections, questions and discussion with students, analysis of CURE assignments regarding what students did and did not understand or accomplish, and course evaluations can also be valuable sources of information. Confusing sections of the lab manual can be better explained, difficult concepts can be addressed more thoroughly in lecture, and limitations or awkward aspects of the tools may be avoided (or even fixed by the developers based on feedback).

%


\section{Results and Discussion}\label{sec12}

\begin{sidewaystable}[h]
\caption{Details of computational CUREs from author's work}\label{tab2}%
\begin{tabular}{llllllll}
\toprule
Semester & Course  & Duration & Topic & Students Involved & No. of    & No. of     & Rough \\
         &         &          &       &                   & Students  & Structures & Runtime \\
\midrule
Fall 2017 & CMP\footnotemark[1] & final project & high-pressure Si phases & PhD: physics & 5 & 5 & 1 hour \\
Spring 2020 & Advanced CMP & final project & Raman of TMDs & PhD: physics, MBSE\footnotemark[2] & 5 & 6 & 1 hour \\
Fall 2021 & CMP & final project & TMD alloys & BS: physics, PhD: physics, MBSE & 17 & 22 & 12 hours \\
Fall 2023 & CMP & final project & TMD alloys & BS: physics & 5 & 22\footnotemark[3] & 12 hours \\
Fall 2022 & Modern Physics & lab session & TMD heterojunctions & BS: physics & 26 & 6 & 1 hour \\
Fall 2023 & Modern Physics & lab session & TMD heterojunctions   & BS: physics & 17 & 8 & 1 hour \\
\botrule
\end{tabular}
\footnotetext[1]{CMP = Condensed Matter Physics}
\footnotetext[2]{MBSE = Materials and Biomaterials Science and Engineering}
\footnotetext[3]{Same set as in fall 2022, with students repeating calculations with anomalous results from previous year.}
\end{sidewaystable}

I have used the paradigm described here for four projects at the University of California, Merced (summarized in Table \ref{tab2}), beginning with graduate courses and refining the ideas over time. In fall 2017, I designed for my graduate Advanced Condensed Matter Physics class a final project on simulation of elastic properties of high-pressure phases of Si, using DFT and MD with Tersoff potentials in nanoHUB, and group theory to constrain the shape of the elastic tensor. In spring 2020, I designed for graduate Advanced Condensed Matter Physics a final project on strain effects on Raman spectra in 2H transition-metal dichalcogenides (TMDs), performed directly with Quantum ESPRESSO on a cluster. This work was presented at a conference \cite{arabi2022}, and a paper is in preparation. In fall 2021 and 2023, I ran for my undergraduate/graduate Condensed Matter Physics class a final project on the energy, structure, and Raman spectroscopy of 2D monolayer MoS$_{2x}$Se$_{2(1-x)}$ alloys. Finally in fall 2022 and fall 2023, I ran for sophomore-level Modern Physics a lab session on in-plane heterostructures of 2H-TMDs \cite{lab_manual}, studied as examples of the particle in a box. These projects will be described in more detail in forthcoming work, regarding both pedagogy and scientific results. The final projects took a few weeks and the lab sessions took about 6 hours, with students working in pairs. %
Assessment so far indicates that students generally found the CURE activities to be interesting and inspiring, and further assessment is ongoing.

\section{Conclusion}\label{sec13}

This article has proposed a computational paradigm for CUREs in materials science and engineering, aiming to inspire instructors elsewhere in this field (and related areas of physics and chemistry) to incorporate CUREs into their course and curricula. A key enabling feature, to keep the focus on materials rather than on details of theory and computational methods, is the use of simplified GUIs from nanoHUB.org that can run research-grade codes via a browser on remote clusters without need for significant computational skills. A notable example is the MIT Atomic-Scale Modeling Toolkit which provides DFT and MD calculations. The calculations can be accessible for undergraduates, for both sophomore-level and upper-division classes. CUREs have positive effects on student attitudes and student learning, and help make undergraduate research more accessible and inclusive. From the instructor's point of view, other benefits include using teaching time and effort to advance their research as an integrated activity, and the ability to use CUREs as an educational and broader impacts activity for grant proposals. Extensive guides on best practices in CUREs, as well as webinar content and teaching materials about this nanoHUB tool, are available to help design CUREs. I encourage instructors to adopt and adapt these ideas to bring CUREs into the field of materials science and engineering.

\backmatter

\bmhead{Acknowledgements}
I thank the students who have taken part in these CURE projects; Marcos Garc\'{i}a-Ojeda and Anubhav Jain for valuable discussions in formulating these ideas; and Enrique Guerrero, Elif Ertekin, Jeffrey C. Grossman, Daniel Richards, and Justin Riley for their contributions to the development of the MIT Atomic-Scale Modeling Toolkit. The nanoHUB team, especially Steven Clark, provided essential technical support. This material is based upon work supported by the National Science Foundation under Grant No. DMR-2144317 and by Cottrell Scholar award No. 26921, a program of Research Corporation for Science Advancement.

\section*{Data availability}

The author declares that the data supporting the findings of this study are available within the paper.

\section*{Conflict of interest}

On behalf of all authors, the corresponding author states that there is no conflict of interest.

\bibliography{sn-bibliography}%

\section*{Author contributions}
D. A. S. designed the educational activities and wrote the article.

\section*{Funding}
National Science Foundation Grant No. DMR-2144317, Cottrell Scholar award No. 26921 from the Research Corporation for Science Advancement.
\end{document}